# Magnetic Properties of Superconducting Cobalt Oxides $Na_xCoO_2 \cdot yH_2O$

Yoshiaki KOBAYASHI, Taketo MOYOSHI, Hidekazu WATANABE, Mai YOKOI and Masatoshi SATO*

*Department of Physics, Division of Material Science, Nagoya University, Furo-cho, Chikusa-ku, Nagoya 464-8602*

Studies of the NMR Knight shift $K$ of $Na_{0.3}CoO_2 \cdot yH_2O$ have been carried out in detail. The suppression of $K$ by the occurrence of the superconductivity reported previously by the present authors in both magnetic field directions perpendicular and parallel to the $c$ axis has been confirmed, indicating that the Cooper pairs are in the singlet state. The anisotropy of the suppression amplitudes is consistent with the anisotropy of the hyperfine coupling constant $A_{spin}$ estimated by the $K$-$\chi$ plot. It has also been found that even samples, which do not exhibit a significant amount of the Curie-Weiss-like increase of the uniform magnetic susceptibility $\chi$ with decreasing temperature $T$, exhibit the superconducting transition, which indicates that the superconducting $Na_{0.3}CoO_2 \cdot 1.3H_2O$ is not necessarily be in the proximity region of the ferromagnetic phase. It has also been confirmed that the superconducting transition temperature $T_c$ of the samples prepared by mixing $Na_{0.7}CoO_2$, $H_2O$ (or $D_2O$) and bromine for 4 h, and separating the resultant powder from the solution by filtration, depends on the elapsed time $t$ after filtration. For these samples, the Curie constant of $\chi$ estimated at low temperatures increases with $t$, supporting the idea that the number of the lattice imperfection possibly due to the oxygen-vacancy-formation increases with $t$. Even for such samples, the superconductivity appears, which seems to exclude the possibility of the anisotropic superconducting order parameter. These results are favorable for the full-gapped superconductivity.

## 1. Introduction

The superconductivity found in $Na_{0.3}CoO_2 \cdot 1.3H_2O$[1] with the superconducting transition temperature $T_c$~4.5 K has attracted much interest because strong magnetic fluctuations inherent in the strongly correlated electrons of the triangular lattice of $CoO_2$ planes, may exist in the system and an unconventional (exotic) superconductivity is realized by the fluctuations. Spin state transitions often observed in metallic Co oxides with varying temperature $T$ or with changing the chemical composition[2-4] also suggests that magnetically active metallic state is easily realized in Co oxides.

Investigations to clarify if the superconducting state of $Na_{0.3}CoO_2 \cdot 1.3H_2O$ is exotic or not, have been carried out from both theoretical and experimental sides, and the possibility of the exotic nature has been pointed out by many authors.[5-10] Meanwhile, results of the $^{59}$Co-NMR Knight shift measured by the present authors' group for both magnetic field directions parallel and perpendicular to the $CoO_2$ planes show that the spin susceptibility $\chi_{spin}$ is suppressed in the superconducting state,[11,12] indicating that the Cooper pairs are in the singlet state. We have also reported that the doping effect of non-magnetic impurities on $T_c$ is too small to be explained by the pair breaking of the impurities for the anisotropic order parameter.[13] Moreover, inelastic neutron scattering study on large single crystals of $Na_{0.3}CoO_2 \cdot 1.3D_2O$ has not detected signals of magnetic fluctuations.[14] These results obtained by the present authors' group contradict to the idea of the exotic superconductivity, but seem to be consistent with the phonon-mediated superconductivity proposed by Yada and Kontani.[15]

In the present work, we have carried out further studies on the magnetic properties of $Na_{0.3}CoO_2 \cdot 1.3H_2O$ specimens prepared with various different conditions to clarify if the magnetism, which is often discussed to be closely related to the occurrence of the superconductivity,[16,17)] really plays a positive role in realizing the superconductivity. In particular, physical properties of the specimens have been studied as functions of time $t$ elapsed after the filtration from the aqueous bromine solution used for the $Na^+$ de-intercalation, where we have found that the $t$-dependence of $T_c$ reported by Barnes *et al.* [18)] is basically reproduced. Based on results of these detailed studies, we argue the mechanism of the Cooper pairing in $Na_{0.3}CoO_2 \cdot 1.3H_2O$.

*) corresponding author (e-mail: e43247a@nucc.cc.nagoya-u.ac.jp)

## 2. Experiments

Powder samples and single-crystals of $Na_{0.3}CoO_2$ were prepared as described in refs. 12 and references therein. Crystals of $Na_{0.72}CoO_2$ grown by the floating zone (FZ) method were crashed into platelets, and then the Na de-intercalation and the $H_2O$ intercalation were carried out to obtain $Na_{0.3}CoO_2 \cdot 1.3H_2O$. Then, their $c$-axes were aligned.

In our previous report,[12)] we showed results of the Knight-shift measurements for samples, which contain, besides the superconducting crystals (platelets) with a $H_2O$ bi-layer between the $CoO_2$ planes, non-superconducting ones with a $H_2O$ mono-layer between the $CoO_2$ planes (monolayer phase) and non-hydrated ones. (In the powder samples, for which the Knight shift was studied under the external field $\boldsymbol{H}$ within the $CoO_2$ planes, we have detected, in the X-ray and/or NMR measurements, neither the monolayer phase nor the non-hydrated system.) In the present work, we have carried out Knight shift studies on an aligned-crystal(platelet)-sample (sample #3), which does not have an indication of non-superconducting phases in the NMR spectra. It enables us, as described later, to ensure the suppression of the spin susceptibility in the superconducting state for $\boldsymbol{H}$ perpendicular to the planes.

We have also prepared polycrystalline samples by immersing $Na_{0.3}CoO_2 \cdot 1.3H_2O$ in aqueous solution of HCl or NaOH with different pH values. For these samples, the magnetic susceptibilities $\chi$ were measured and the results are used to argue the relationship between the magnetic properties and the superconductivity. The $T_c$ values of the samples prepared, as reported in ref. 18, by mixing $Na_{0.7}CoO_2$, $H_2O$ (or $D_2O$) and bromine for 4 h and then, separating the resultant powder from the solution by filtration, were measured at various time $t$ elapsed after the filtration.

The magnetic susceptibilities were measured using a Quantum Design SQUID magnetometer. The measurements of NMR and NQR were carried out by a standard coherent pulsed NMR method. The NQR (and NMR) spectra were taken by recording the spin-echo intensity with the applied frequency (or field) being changed stepwise. In the measurements of the Knight shift of the aligned-crystal-samples, the central transition ($-1/2 \leftrightarrow 1/2$) line was obtained by the Fourier transform (FT) NMR technique.

## 3. Experimental Results and Discussion

FT-spectra of the central transition lines of $^{59}$Co-NMR obtained for the aligned-crystal-samples #1-#3 with the applied field $\boldsymbol{H}//c$ are shown in the upper panels of Figs. 1(a)-1(c), respectively. The data obtained for $^{23}$Na are also shown for samples #1and #3 in the lower panels of Figs. 1(a) and 1(c), respectively. (The $T_c$ values of the samples #1-#3 are 4.0 K, 4.5 K and 4.0 K, respectively.) The peak positions of the $^{59}$Co-NMR profiles shift toward lower frequency with decreasing $T$ below $T_c$, while the peak positions of the $^{23}$Na-NMR-spectra are nearly $T$-independent even below $T_c$. These facts indicate that the effect of the superconducting diamagnetism $M_{dia}$ is negligibly small, and that the peak shifts observed for the $^{59}$Co-NMR spectra below $T_c$ is due to the suppression of the electronic spin susceptibilities $\chi_{spin}$ in the superconducting phase.

Fig. 2(a) shows the $T$-dependence of the peak frequency of the $^{59}$Co-NMR profile $^{59}f_{peak,\,c}$ obtained for the sample #3 with the applied field $\boldsymbol{H}//c$. In Fig. 2(b) and in the lower panel of Fig. 1(b), the shifts obtained for the samples #3 and #2 are shown in the form of $\Delta^{59}K_c(T) \equiv [^{59}f_{peak,\,c}(T) - {}^{59}f_{peak,\,c}(\sim 4\ \mathrm{K})]/\gamma_N H$ *versus* $T$, respectively, where $\gamma_N$ is the nuclear gyromagnetic ratio (10.03 MHz/T for $^{59}$Co). The $T$-dependence of $\Delta^{59}f_{peak,\,c}$ [$\equiv {}^{59}f_{peak,\,c}(T) -$

$^{59}f_{\mathrm{peak},\,c}$(~4 K)] obtained for the sample #2 is shown in the inset of the lower panel of Fig. 1(b), where the $|\Delta^{59}f_{\mathrm{peak}}|$ at 1.5 K is found to increase with $H$, indicating that the nonzero $|\Delta^{59}f_{\mathrm{peak},\,c}|$ below $T_\mathrm{c}$ is not due to the superconducting diamagnetic moment $M_{\mathrm{dia}}$ but due to the reduction of $\chi_{\mathrm{spin}}$ in the superconducting state.[12] As can be clearly seen in Figs. 1(b) and 2(a), $\Delta^{59}K_c(T)$ or $^{59}f_{\mathrm{peak},\,c}$ gradually increases with $T$ in the region above $T_c$, indicating that the system may exhibit the pseudo-gap behavior.

The $|\Delta^{59}K_c$(~1.5 K)$|$ values of samples #1-#3 are estimated to be 0.03±0.005 %, 0.03±0.005 % and 0.08±0.02 %. The small values of the $|\Delta^{59}K_c$(~1.5 K)$|$ observed for the former two are partly due to the existence of the non-superconducting monolayer parts in these crystals. As we reported previously,[12] the profiles of the satellite lines of $^{59}$Co-NMR spectra often exhibit the double peak structures, which originate from the coexistence of the non-superconducting monolayer phase. Since a resonance frequency $^{59}f_{\mathrm{peak},\,c}$ is estimated in our study from the peak position of the centerline of the spectra, the shift of $^{59}f_{\mathrm{peak},\,c}$ is expected to be smaller than the one for the single phase sample of the superconducting bi-layer system.[12] Fortunately, the satellite-lines of the sample #3 does not, as shown in Fig. 2(c), have such the double peak structure, indicating that the non-superconducting monolayer phase does not coexist, which is in contrast with the cases of samples #1 and #2.[12] The larger value of $|\Delta^{59}K$(~1.5K)$|$ of the sample #3 than the other two can be naturally understood by this consideration. Recently, results of the $\Delta^{59}K$ measurements carried out for aligned powder samples with the applied fields $\boldsymbol{H}$ parallel and perpendicular to $CoO_2$ planes have been reported.[20] Their result for $\boldsymbol{H}$//$CoO_2$ planes is almost identical to our reported one. For $\boldsymbol{H}$//$c$, the error bars of their data are too large to judge if their result agrees with ours or not. Anyhow, we compare their data with ours in Fig. 3.

Figure 4(a) shows the uniform magnetic susceptibilities $\chi$ of a single crystal of $Na_{0.35}CoO_2{\cdot}1.3H_2O$ (sample #4) measured with $H$=1 T applied along two different directions. Figures 4(b) and 4(c) show the $T$-dependence of Meissner (or shielding) signals for $H$=10 G applied parallel and perpendicular to the $c$ axis. We note here that the superconductivity occurs even in the sample that does not exhibit a significant $\chi$-increase with decreasing $T$ at low temperatures. Similar $\chi$-$T$ curves of $Na_{0.35}CoO_2{\cdot}1.3H_2O$ was also reported by Jin *et al.*[21] Then, a large increase of $\chi$ with decreasing $T$ below ~100 K in the $\chi$-$T$ curves reported previously[12,19] is not intrinsic, or it is not important for the occurrence of the superconductivity. The idea is confirmed by the almost $T$-independent behavior of the $^{59}$Co-Knight shifts $^{59}K_{ab}$ at low temperatures shown in the inset of Fig. 3 of ref. 12 (see also the lower panel of Figs. 1(b) for $^{59}K_c$). Ning *et al.*[22] and Mukhamedshin *et al.*[23] have also reported that the $T$ dependence of $K$ below ~100 K is very small. Since the difference of the susceptibilities, $\Delta\chi(T)$ between $\chi_{ab}$ for $\boldsymbol{H}$ // $ab$-plane and $\chi_c$ for $\boldsymbol{H}$ // $c$ is almost $T$-independent (see Fig. 4(a)), $\chi_{\mathrm{spin}}$ is nearly isotropic and the difference can be considered to be just due to the anisotropic orbital (Van Vleck) susceptibility. This contradict the results reported by Chou *et al.*[24]

Figure 5(a) shows the $K_c$-$\chi_c$ plots obtained for the sample #3. Using this, the values of $K_{\mathrm{spin}}{}^{c}$ at ~5 K (>$T_\mathrm{c}$) and the hyperfine coupling constant $A_{\mathrm{hf}}{}^{c}$ for $\boldsymbol{H}$//$c$ are estimated to be ~0.25 % and 40 ± 5 kOe/$\mu_\mathrm{B}$, respectively. Comparing these values with $K_{\mathrm{spin}}{}^{ab}$ at ~5 K and $A_{\mathrm{hf}}{}^{ab}$ (~0.45 % and 88 ± 2 kOe/$\mu_\mathrm{B}$, respectively[12]), we find that $A_{\mathrm{hf}}{}^{c}$ is roughly a half of $A_{\mathrm{hf}}{}^{ab}$. Then, we draw in Fig. 5(b) the thick dotted line as the $\Delta K_c$-$T$ curve expected for purely superconducting parts. (Note that the $\Delta K_c$ value is about a half of the shift $\Delta K_{ab}(T)$ shown by the broken line. As for the anisotropic hyperfine coupling, the dipolar field at the $^{59}$Co sites may be responsible.) Because the sample #3 does not show, as we stated above, any experimental indications of the non-superconducting monolayer parts, the curve can be compared with the experimentally observed data (squares), if only the consideration of possible effects of normal vortex cores is made. We think that the $|\Delta K_c|$ value is large enough to indicate that the spin susceptibility is almost completely suppressed even in the case $\boldsymbol{H}$//$c$.

For powder samples of $Na_xCoO_2{\cdot}yH_2O$ prepared by immersing powders of $Na_{0.3}CoO_2{\cdot}1.3H_2O$ in HCl (or NaOH) aqueous solutions with various values of pH, the lattice parameters $a$ and $c$ are shown in Fig. 6(a) against the pH value of the solution: The parameter $a$ increases slightly and $c$ decreases from ~19.8 to ~19.7 Å with increasing pH in the region of pH <7 and seems not to change in the region pH >7. The similar behavior of $c$ has been reported in ref. 17. In the present study, we have found that only the samples of $Na_xCoO_2 \cdot yH_2O$ immersed in aqueous solutions with the pH values of about 2, 3.5, 7.5 and 12.5 exhibit superconductivity with the $T_c$ values of about 4.0, 2.9, 4.5 and 4.5 K, respectively. The others do not, down to 1.9K. Figure 6(b) shows the $T$-dependences of $\chi$ for the superconducting and non-superconducting samples prepared in the solutions with pH~12.5 and ~11, respectively (hereafter, the samples are called samples A and B, respectively. See also the inset of Fig. 6(b) for their Meissner or shielding diamagnetism.). In the figure, we find that their $\chi$-$T$ curves overlap almost completely, indicating that the occurrence of the superconductivity do not correlate with the behavior of the uniform magnetic susceptibility.

In Figs. 7(a) and 7(b), the lattice parameters $a$ and $c$ and the transition temperature $T_c$ determined from the superconducting diamagnetism, respectively, are shown against the time $t$ elapsed after the filtration of the samples from the admixture of bromide and water. The parameter $a$ does not sensitively depend on $t$, while $c$ decreases linearly in $t$. $T_c$ exhibits an initial increase with $t$, have the maximum (~4.7 K) at $t$ = 175 h, and decreases drastically with the further increase of $t$. We could not observe the superconducting transition down to 1.9 K for $t \geq 230$ h. These results are very similar to those reported by Barnes *et al.*[18]

The $T$ dependences of $\chi$ measured at several $t$ values after filtration are shown in Fig. 8(a). Below ~50 K, $\chi$ increases with $t$. We have just tried to fit the simple function $\chi = C/(T+\theta) + \chi_0$ to the observed data in the $T$ region lower than ~15 K, where $C$ is Curie constant and $\chi_0$ is the $T$-independent terms, and the results are shown in Figs. 8(b) and 8(c): $C$ increases with $t$, $\theta$ is almost $t$ independent and $\chi_0$ decreases with $t$. We have found that $\Delta C$ $\{\equiv C(t{=}216\text{ h}) - C(t{=}0)\}$ is 0.012 emu/K·mol, which corresponds to the number of the localized spins (S=1/2), $\Delta N = 0.03N_A$, where $N_A$ is Avogadro's constant. According to the neutron structural analysis of Barnes *et al.*,[18] the number of oxygen vacancies, $\delta$ is estimated to be ~0.1 at $t$~ 216 h, where $T_c$ disappears with increasing $t$. This $\delta$ value corresponds to the cobalt valence of +3.47 (for $Na_xCoO_2 \cdot yH_2O$ with $x$=0.33). If 3 % of the conduction electrons are localized, the effective valence of cobalt ions is estimated to be 3.50. At this valence, the tendency of certain kind of ordering, which is actually observed in $Na_{0.5}CoO_2$, is enhanced, and may be responsible for the rapid disappearance of the superconductivity.

Now, We have shown that the spin susceptibility is definitely suppressed by the superconductivity in both magnetic field directions perpendicular and parallel to the $c$ axis. We have also shown that even the samples which do not have the significant amount of the Curie-Weiss-like term of the magnetic susceptibility, exhibit the superconducting transition. It indicates that the superconducting $Na_{0.3}CoO_2 \cdot 1.3H_2O$ is not necessarily be in the proximity region of the ferromagnetic phase. It has been confirmed that $T_c$ of the samples prepared by mixing $Na_{0.7}CoO_2$, $H_2O$ (or $D_2O$) and bromine for 4 h, and separating the resultant powder from the solution by filtration, depends on the elapsed time $t$ after filtration. For these samples, the Curie constant estimated at low temperatures has been found to increase with $t$, supporting the idea that the number of the lattice imperfection possibly due to the oxygen vacancy formation increases with $t$. Even for such samples, the superconductivity appears, which seems to exclude the possibility of the anisotropic superconducting order parameter. All these results are favorable for the fully-gapped superconductivity.

Acknowledgments –The work is supported by Grants-in-Aid for Scientific Research from the Japan Society for the Promotion of Science (JSPS) and by Grants-in-Aid on priority area from the Ministry of Education, Culture, Sports, Science and Technology.

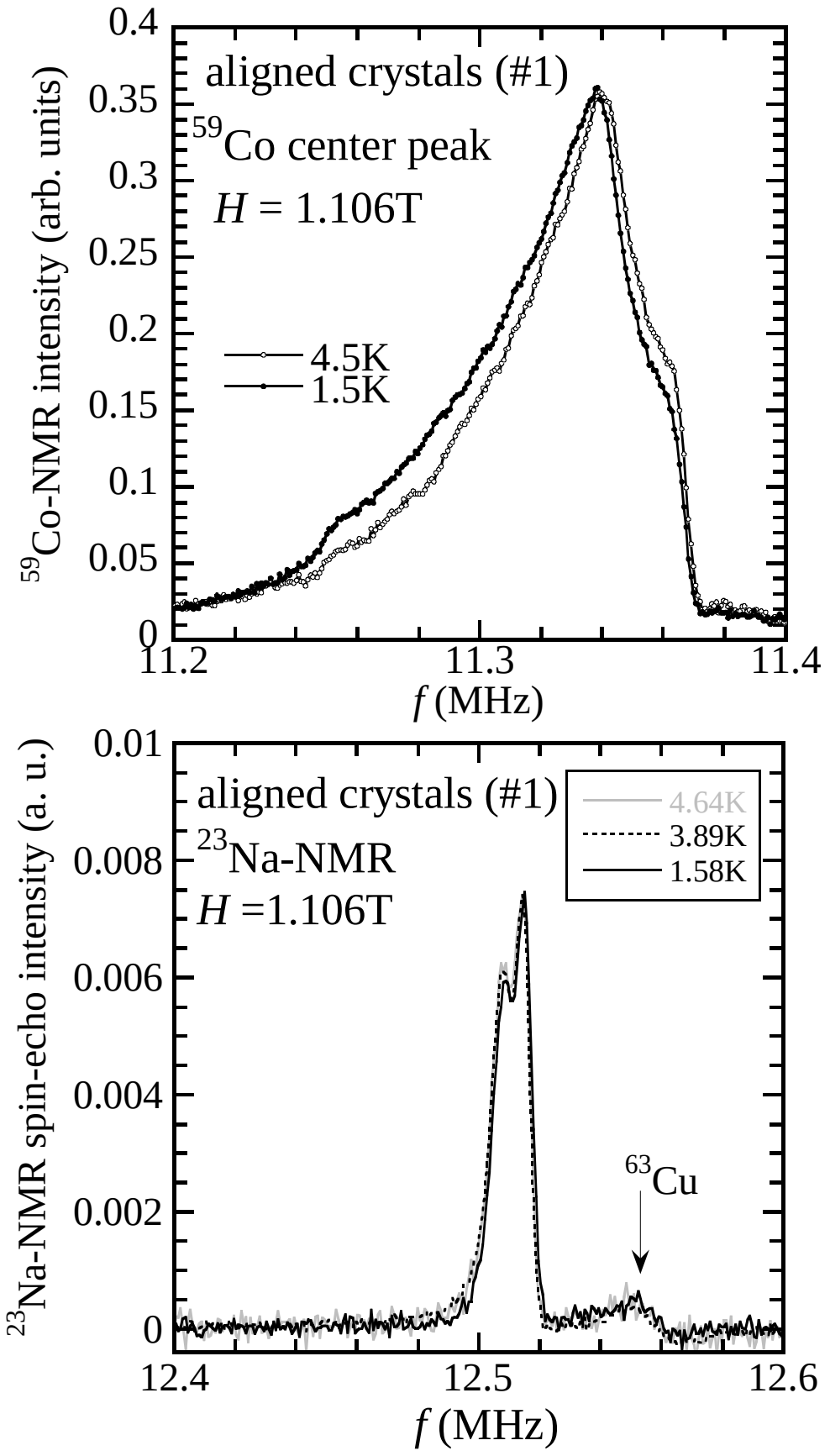

Fig. 1(a) FT-spectra of the center peak of $^{59}$Co-NMR taken at $T$ =4.5 K (>$T_c$) and 1.5 K (<$T_c$) for the aligned-crystal sample #1 with $\boldsymbol{H}\parallel c$ are shown in the upper panel. FT-spectra of center peak of $^{23}$Na-NMR taken for the same sample for $\boldsymbol{H}\parallel c$-axis at $T$ = 4.64, 3.89 and 1.58 K are shown in the lower panel.

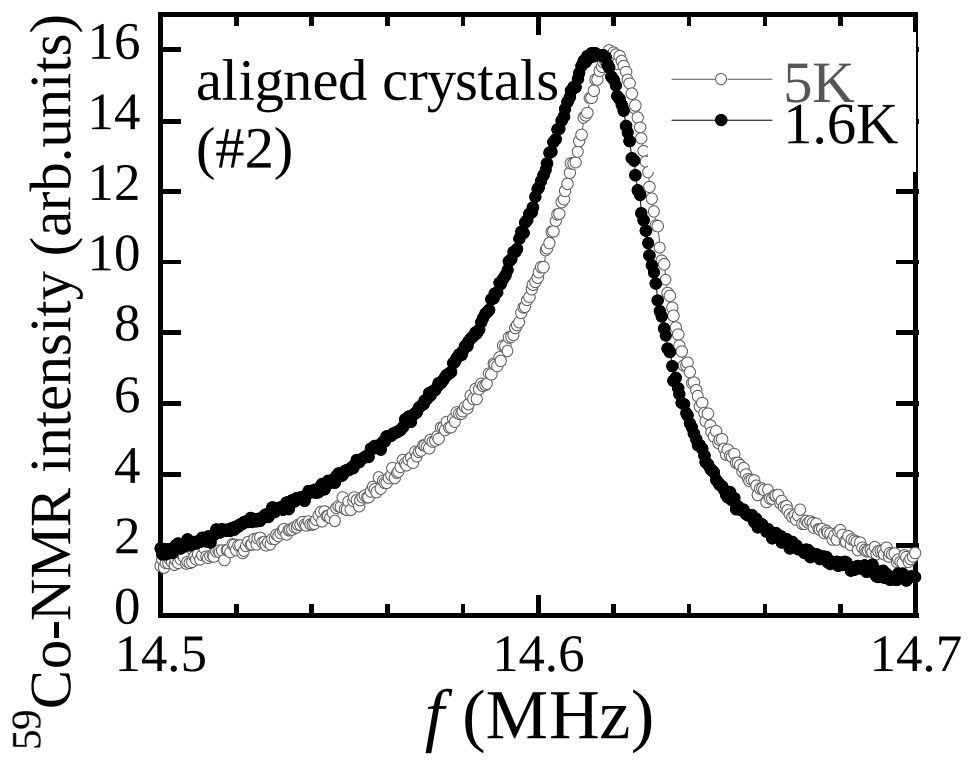

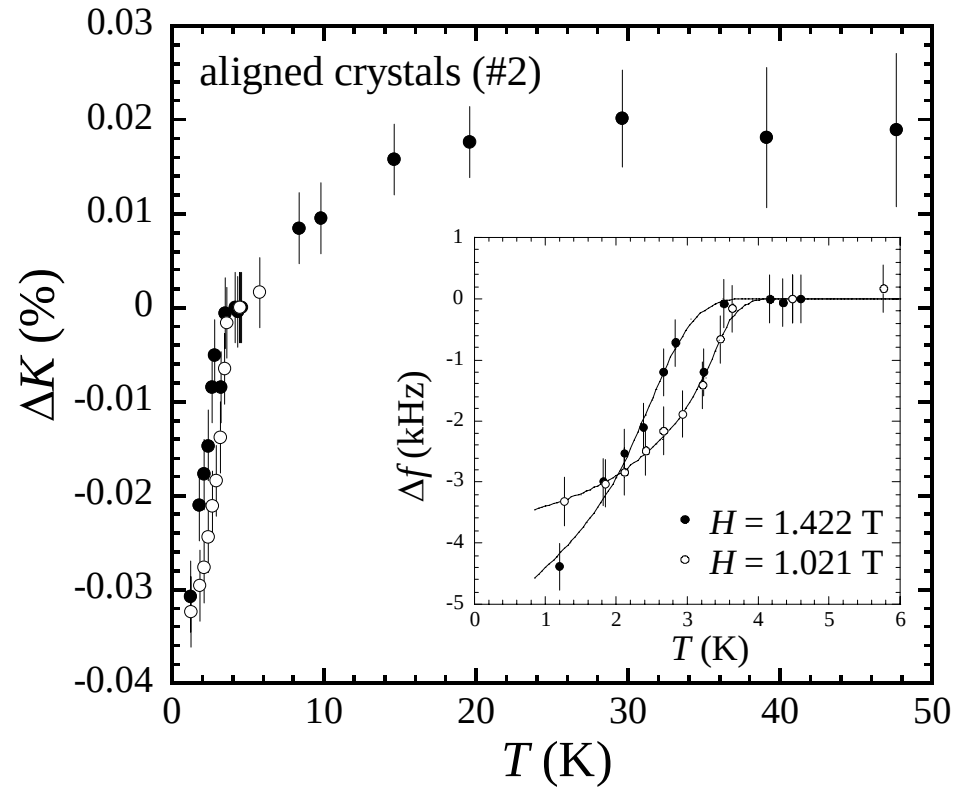

Fig. 1(b) FT-spectra of the center peak of $^{59}$Co-NMR taken at $T$ =5.0 K (>$T_c$) and 1.6 K (<$T_c$) for the aligned-crystal sample #2 with $\boldsymbol{H}\parallel c$ are shown in the upper panel. In the lower panel, $\Delta K(T)$ ($\equiv K(T)-K(4.6\text{K})$) is plotted against $T$ for $^{59}$Co-NMR Knight shifts. Inset shows the $T$-dependence of $\Delta f$ [$=f(T)-f(4.6\text{K})$], where $f$ is resonance frequency estimated from the peak position of NMR spectra.

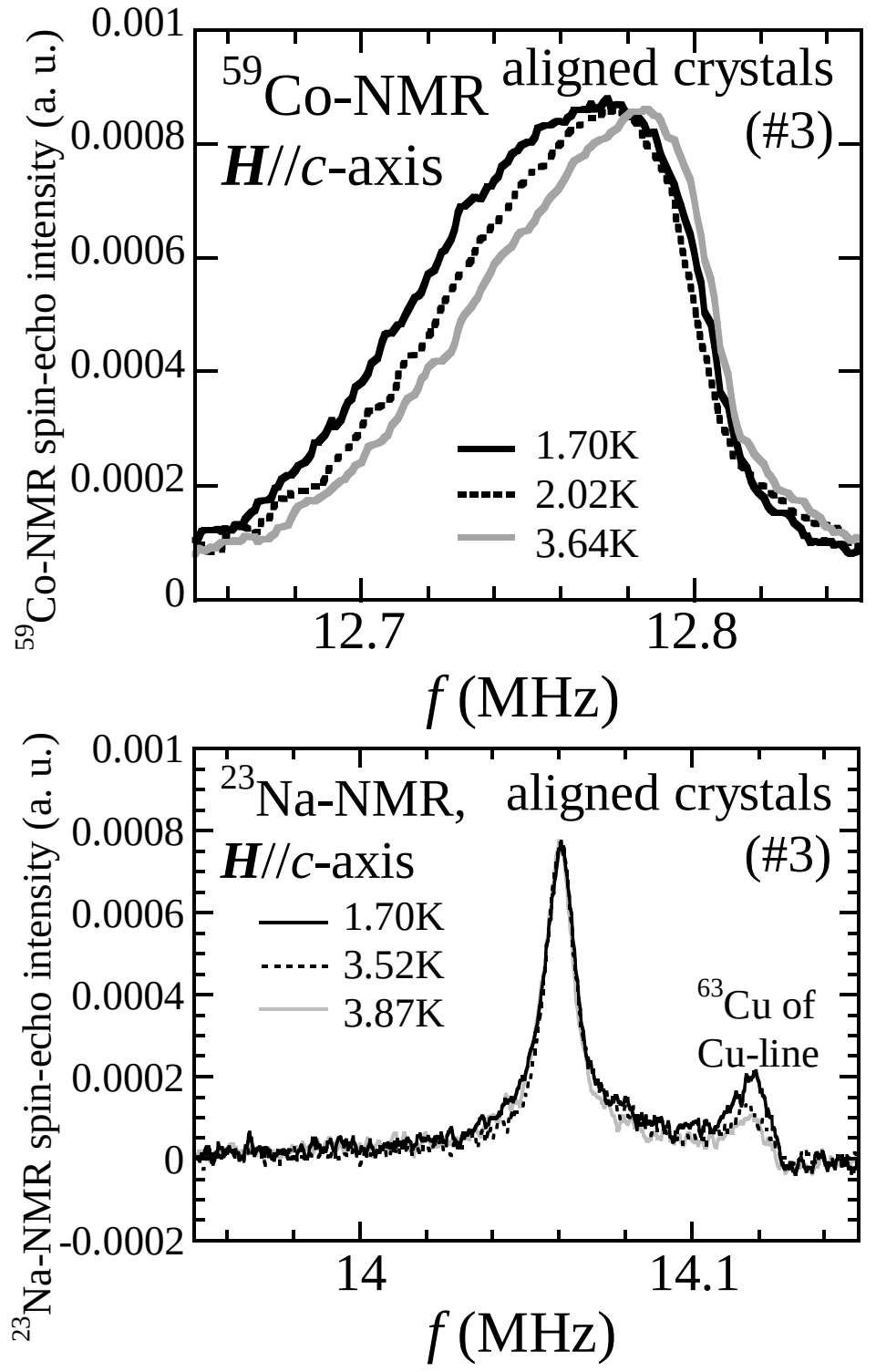

Fig. 1(c) Similar figures to those in Fig. 1(a) are shown for the aligned-crystal sample #3.

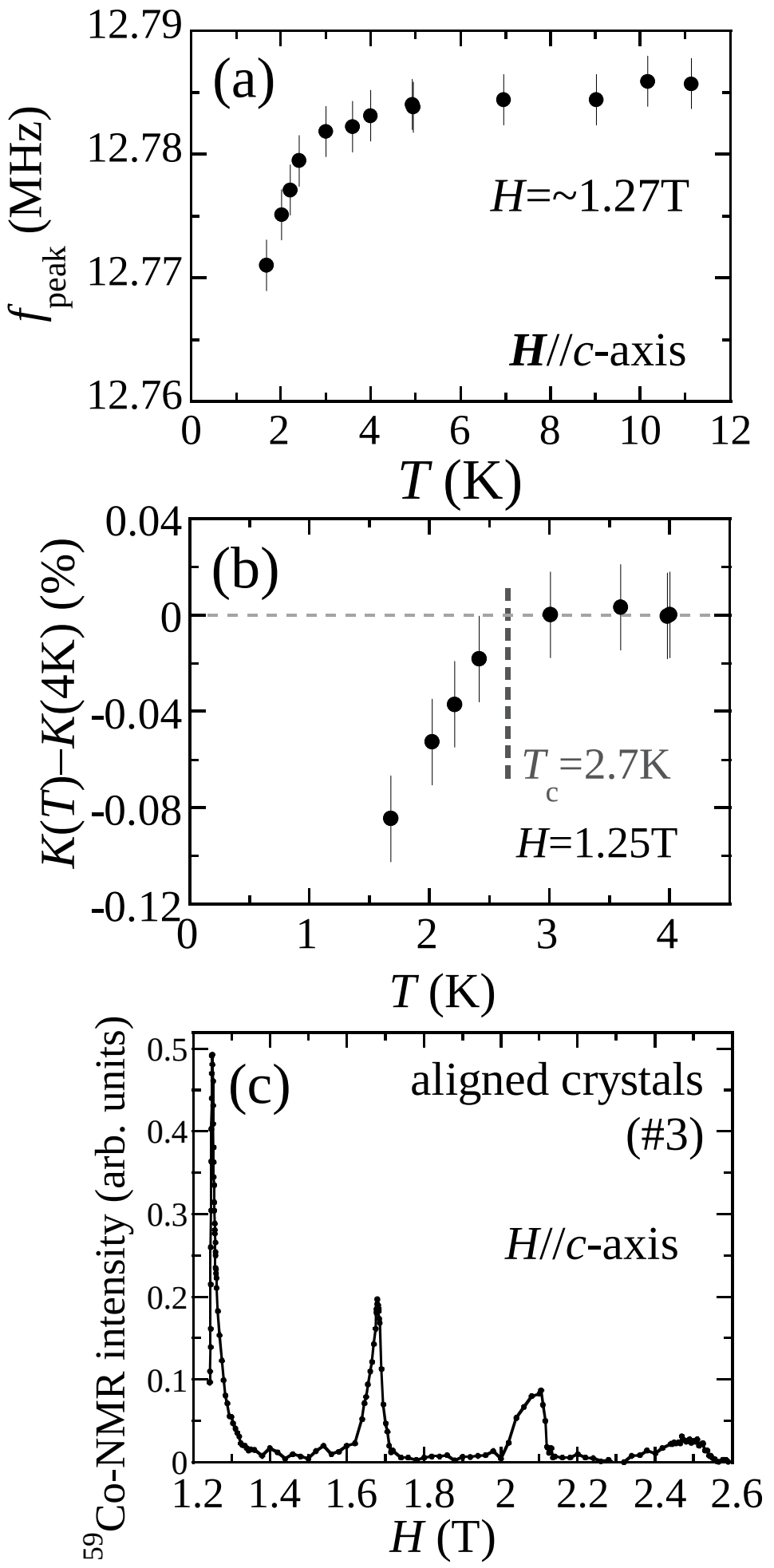

Fig. 2 In (a), the peak frequency $f$ of the FT-spectra of the center peak of $^{59}$Co-NMR taken with $H$~1.27 T is plotted against $T$ for the sample #3. (b) $\Delta K(T)$ ($\equiv K(T)-K$(4.6K)) taken with $H$~1.27 T for the sample #3 is plotted against $T$. (c) $^{59}$Co-NMR intensity observed for the sample #3 is shown as a function of the applied magnetic field $H$.
The satellite lines do not split, indicating that the non-superconducting monolayer par does not have a significant volume fraction in the sample.

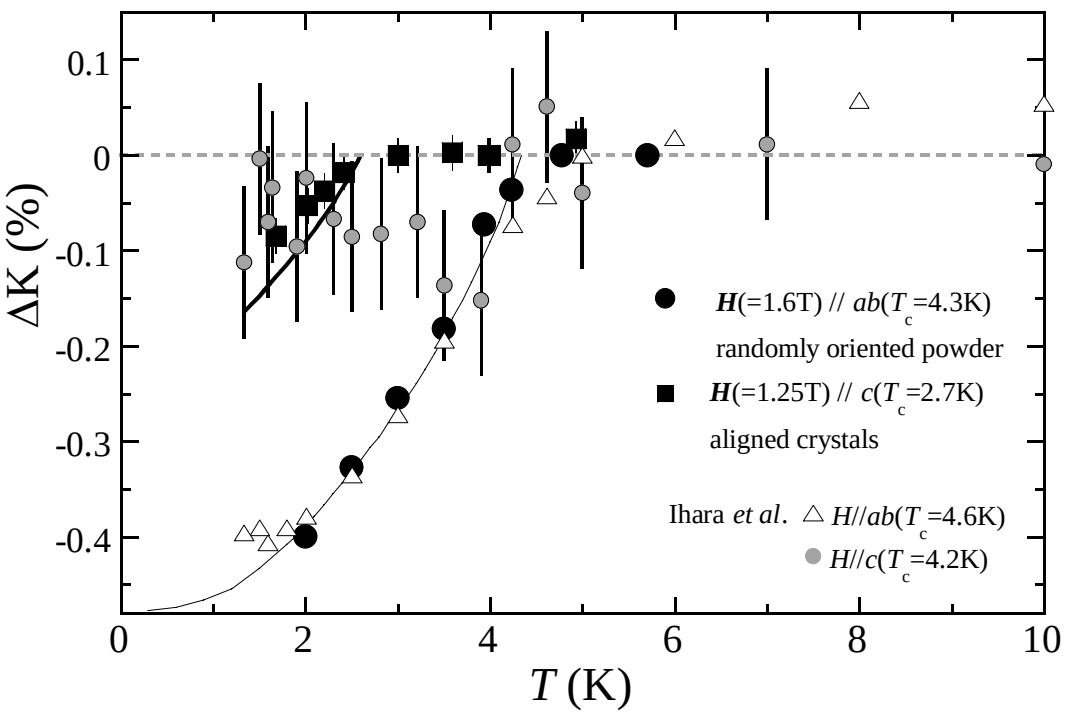

Fig. 3 Comparison of $\Delta K(T)$ of our data with those published by Ihara *et al*.[16] Closed circles are the data for H//($ab$-plane) published by the present authors,[11] and the closed squares are the data for the sample #3 with $H$//$c$, which confirm the suppression of $K$ reported by the present authors, previously.[12] The gray circles and open triangles are the data for $\boldsymbol{H}$ parallel and perpendicular to the $c$-axis recently reported in ref. 16

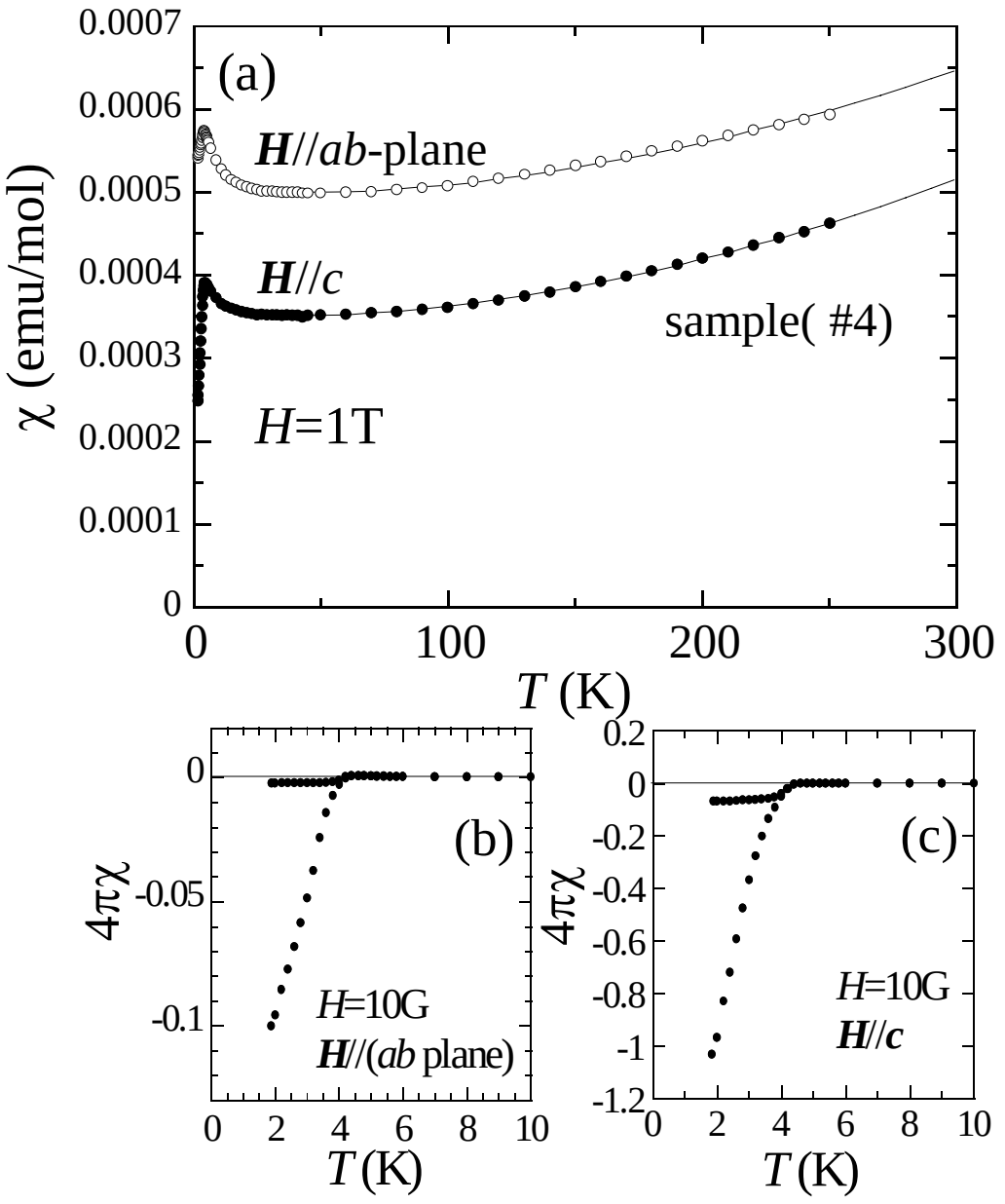

Fig. 4 (a)Magnetic susceptibilities measured for single crystals of $Na_{0.3}CoO_2 \cdot 1.3H_2O$ with $\boldsymbol{H}$ parallel and perpendicular to $c$ are shown against $T$.
(b) and (c) show the superconducting diamagnetism observed under the conditions of the zero-field -cooling and the field-cooling with $\boldsymbol{H}$ perpendicular and parallel to $c$, respectively.

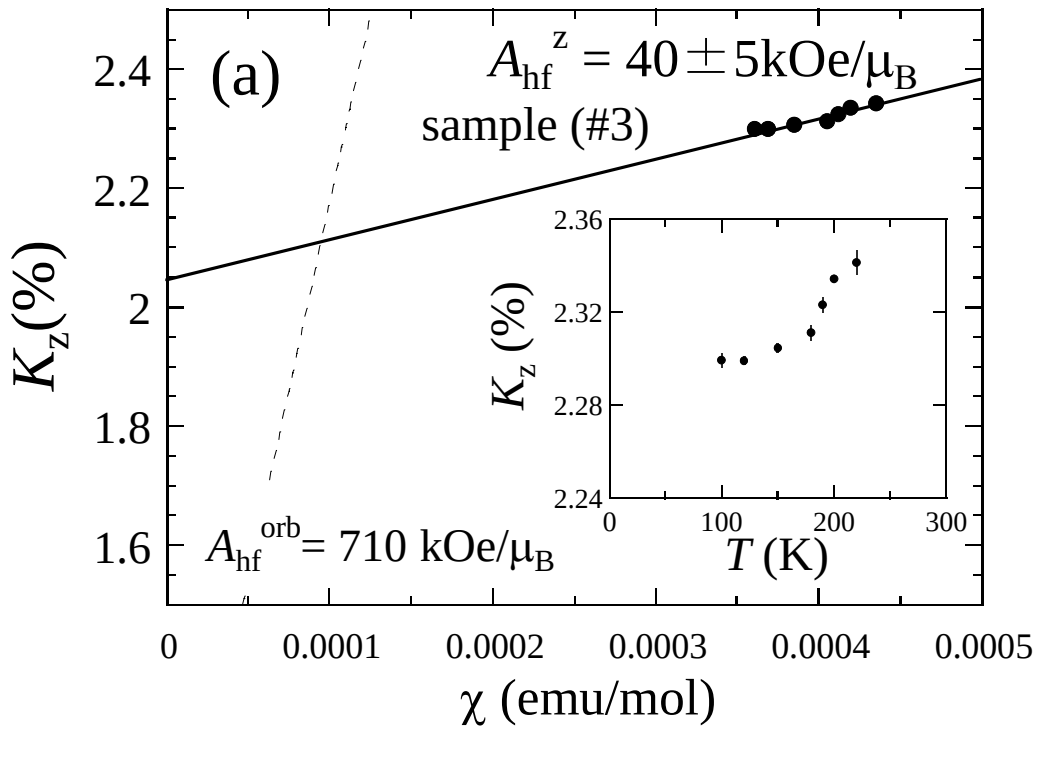

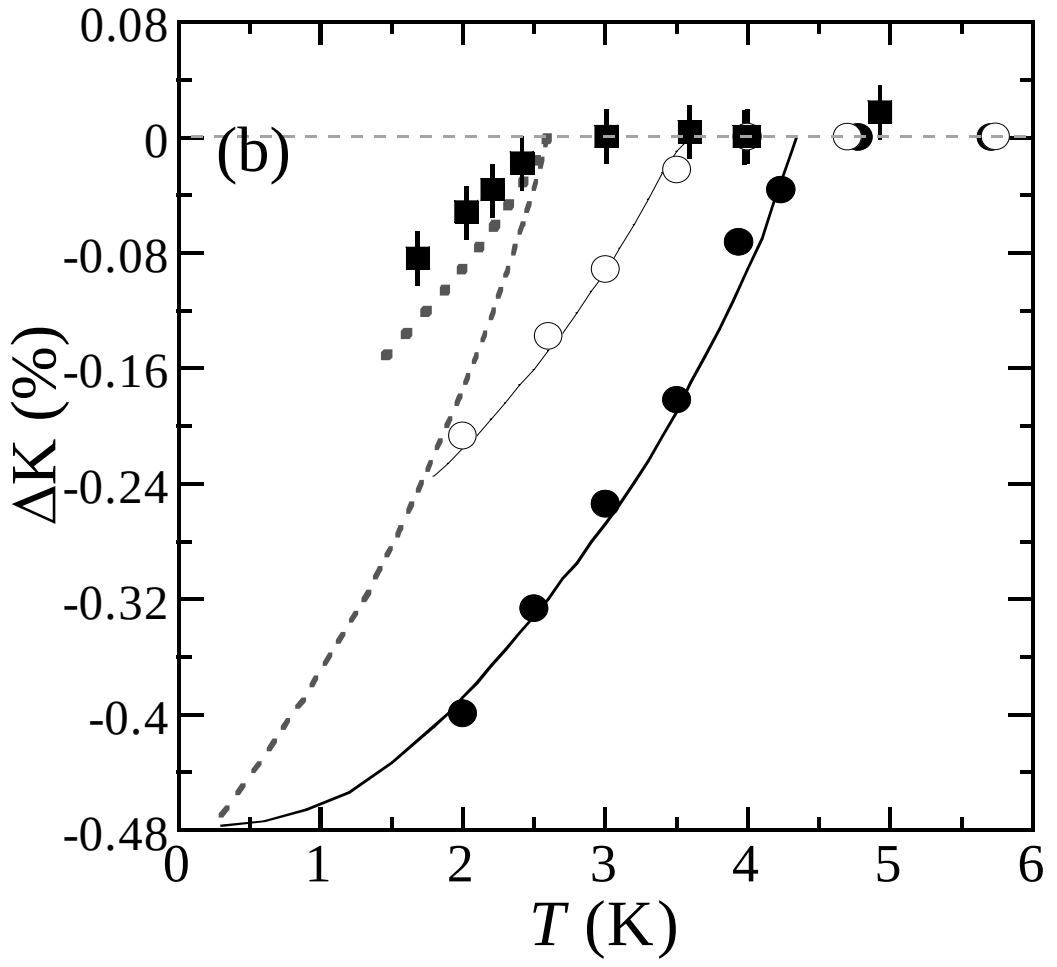

Fig. 5 In (a), the $K$-$\chi$ plot for $^{59}$Co is shown for the sample #3. The inset shows the $T$ dependence of the $^{59}K$ for $\boldsymbol{H}$ parallel to $c$. (b) The $T$ dependences of the observed data of $\Delta K(T)$ ($\equiv K(T)$– ($T>T_c$)) are shown (see also Fig. 2(a)), where the closed and open circles are taken with two different magnitudes of the applied magnetic fields $\boldsymbol{H}$ for a polycrystalline sample by using a peak corresponding to $\boldsymbol{H}$ within *ab* planes,[12] and the closed squares are the data taken for $\boldsymbol{H}$ perpendicular to *ab* planes. The thin and thick broken lines shows the $T$ dependences of $K$ expected for the case $\boldsymbol{H}//c$ in two cases, where the hyperfine coupling constant is equal to that for $\boldsymbol{H}$ within *ab* planes and it has the value estimated by $K$-$\chi$ plot, respectively. See text for details.

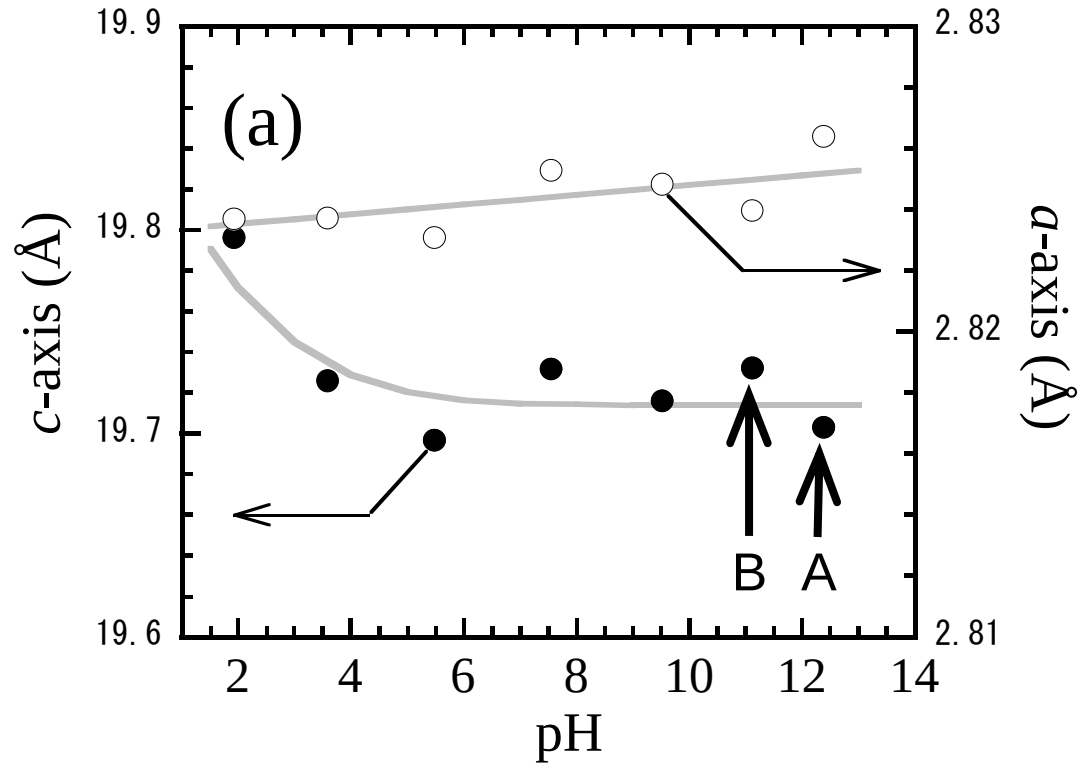

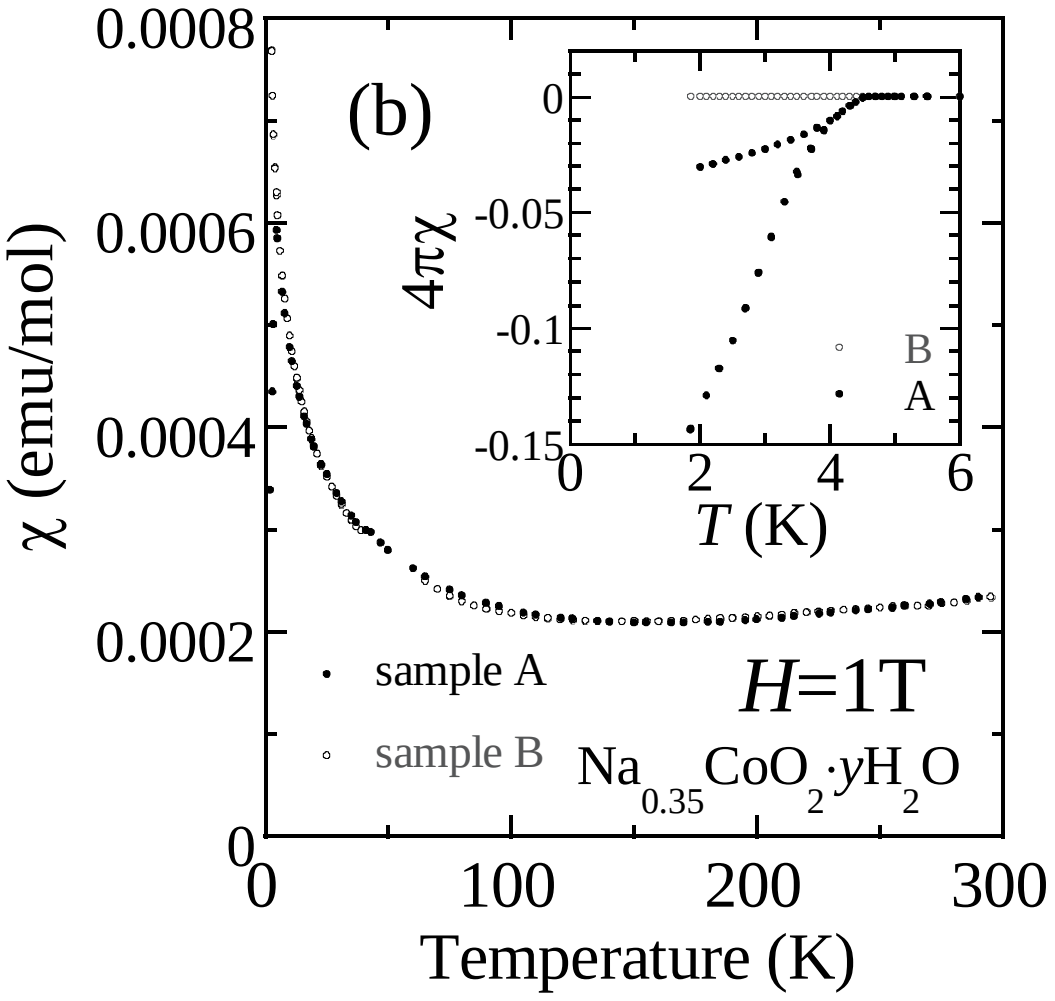

Fig. 6 (a) Lattice parameters of the samples of $Na_xCoO_2 \cdot H_2O$ are shown against the pH value of the HCl (or NaOH) aqueous solution, into which original powders of $Na_xCoO_2 \cdot H_2O$ were immersed to obtain the samples used for the studies of the preparation-condition -dependence. (b) Magnetic susceptibilities $\chi$ of the samples shown in (a) by A (superconducting) and B (non-superconducting at least down to 1.9 K). We can find that their magnetic susceptibilities are almost identical to each other's, suggesting that the occurrence of the superconductivity does not correlate with the behavior of $\chi$. Inset shows the superconducting diamagnetism for the samples.

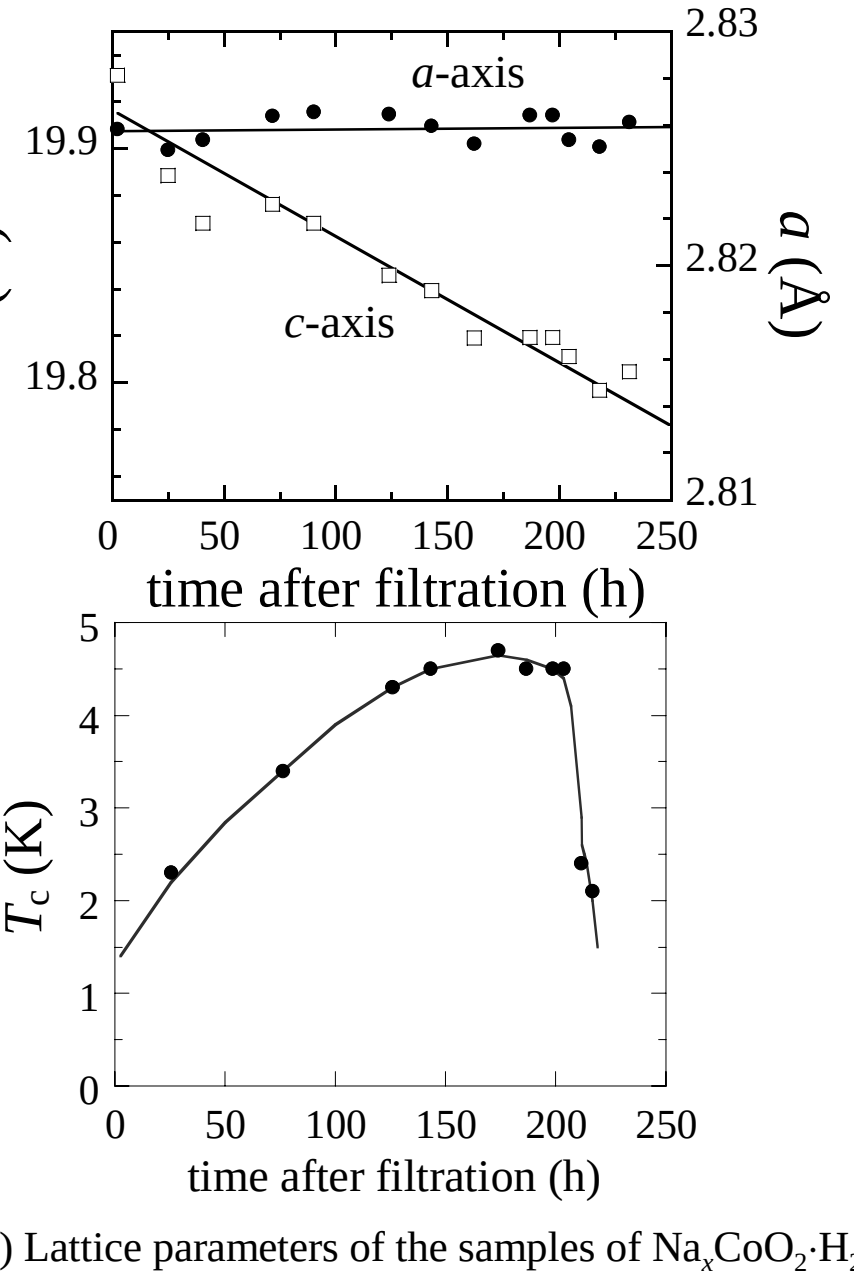

Fig. 7 (a) Lattice parameters of the samples of $Na_xCoO_2 \cdot H_2O$ are shown against the time $t$ elapsed after the filtration of the samples from the admixture of bromide and water. (b) Superconducting transition temperature $T_c$ measured for the samples are shown as a function of the time $t$ elapsed after the filtration.

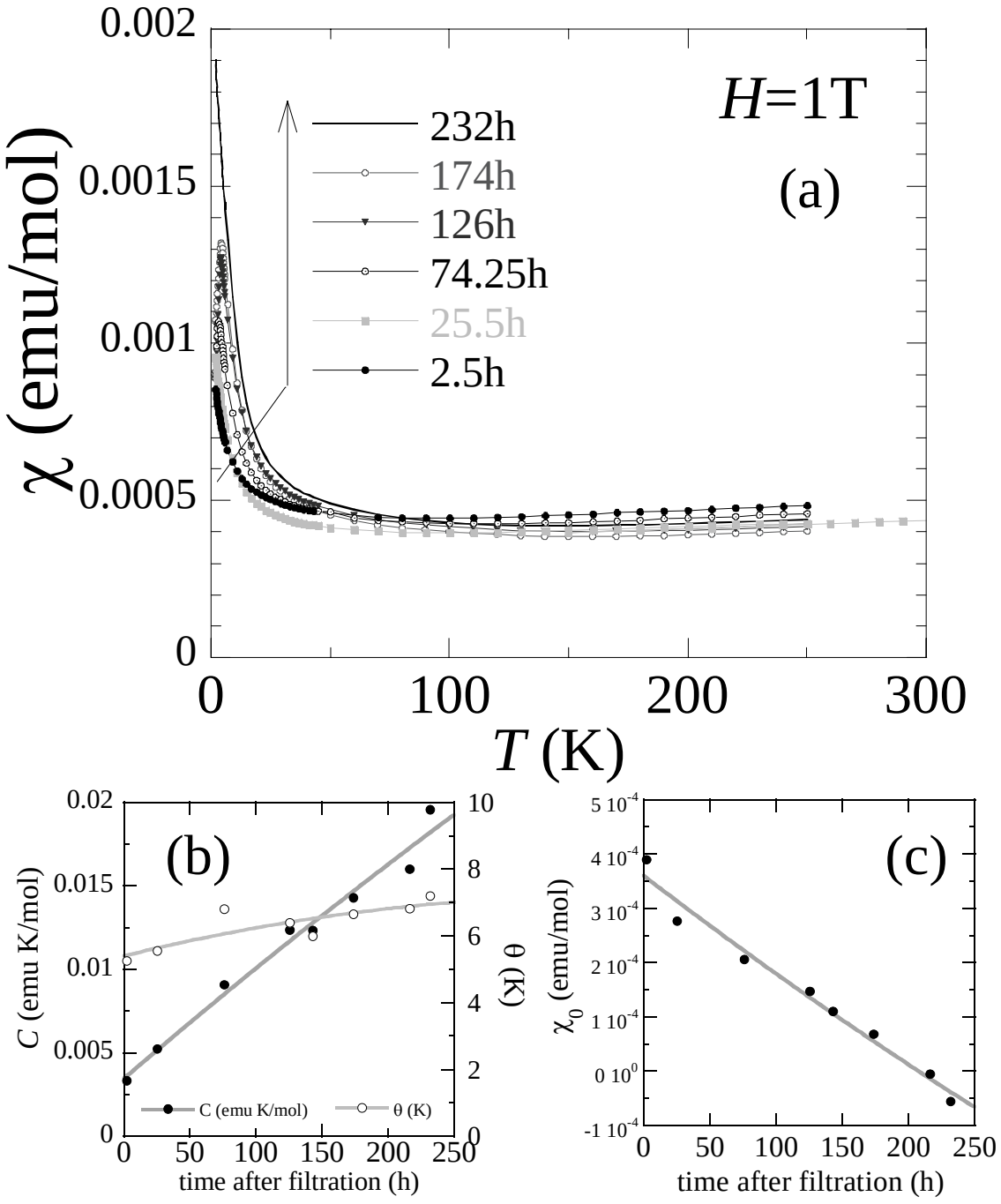

Fig. 8 (a) $T$ dependence of the magnetic susceptibility $\chi$ is shown for the samples obtained by the filtration from the admixture of bromide and water. For the samples, the parameters $C$ and $\theta$, and $\chi_0$ obtained by fitting the simple relation $\chi=\chi_0+C/(T+\theta)$ to the observed data of $\chi$ in the $T$ range below 15 K are shown as functions of the time $t$ in (b) and (c), respectively.